\documentclass[fleqn,usenatbib]{mnras}

\usepackage{newtxtext,newtxmath}

\usepackage[T1]{fontenc}

\DeclareRobustCommand{\VAN}[3]{#2}
\let\VANthebibliography\thebibliography
\def\thebibliography{\DeclareRobustCommand{\VAN}[3]{##3}\VANthebibliography}

\usepackage{graphicx}	
\usepackage{amsmath}	
\usepackage{lscape}
\usepackage{subfig}
\usepackage{makecell}
\usepackage{threeparttable}
\usepackage{hyperref}
\usepackage{url}

\title[Compact star merger with interacting SQM]{Compact star merger events with stars composed of interacting strange quark matter}

\author[A. Kumar, V. B. Thapa, M. Sinha]{
Anil Kumar\thanks{anil.1@iitj.ac.in}, 
Vivek Baruah Thapa\thanks{thapa.1@iitj.ac.in} 
and Monika Sinha
\thanks{ms@iitj.ac.in}
\\
Indian Institute of Technology Jodhpur, Jodhpur 342037, India
}

\pubyear{2022}

\usepackage[normalem]{ulem}  

\begin{document}
\label{firstpage}
\pagerange{\pageref{firstpage}--\pageref{lastpage}}
\maketitle

\begin{abstract}
We investigate the properties of stars participating in double compact star merger events considering interacting model of stable strange quark matter. We model the matter making it compatible with the recent astrophysical observations of compact star mass-radius and gravitational wave events. In this context we consider modified MIT bag model and vector bag model with and without self interaction. We find new upper bound on tidal deformability of $1.4~M_\odot$  strange star corresponding to the upper bound of effective tidal deformability inferred from gravitational wave event. Range of compactness  of $1.4~M_\odot$ strange star is obtained as ${0.175}\leq{C_{1.4}}\leq{0.199}$. Radius range of $1.5M_\odot$ primary star is deduced to be ${10.57}\leq{R_{1.5}}\leq{12.04}$ km, following stringent GW170817 constraints. GW190425 constraints provide with upper limit on radius of $1.7$ solar mass strange star that it should be less than $13.41$ $\text{km}$.
\end{abstract}




\section{Introduction}\label{sec:1}

Massive stars end their lives by supernova explosion to produce very compact objects of average density $\sim 10^{14}$ gm cm$^{-3}$. The intense gravitational fields squeeze the matter present inside these compact objects to densities ranging from sub-saturation to few times nuclear saturation density $n_0=0.16$ fm$^{-3}$ \citep{1996cost.book.....G, Sedrakian2007PrPNP}.  The true nature of matter at such high densities is not well constrained by any terrestrial experiment and hence the internal matter composition and properties of these stars are yet not entirely understood. Only way to understand highly dense matter is the astrophysical observations from these compact objects. The recent detection of gravitational-waves (GWs) has opened up new ventures to explore dense matter behaviour and constrain high density equation of state (EOS). The GW emissions as obsereved by LIGO-Virgo Collaboration, from GW170817 \citep{LIGO_Virgo2017c,LIGO_Virgo2017a, LIGO_Virgo2017b}, GW190425 \citep{2020ApJ...892L...3A} and GW190814 \citep{2020ApJ...896L..44A} events are due to spiraling compact stars made of dense matter scenario. In this scenario, the involved compact objects tidally deform each other emitting GW signal \citep{PhysRevD.81.123016}. Analysis of GW observations reveals that the matter should be soft at low density from the maximum bound of dimensionless tidal deformability $\tilde{\Lambda}$. 
It indicates that the matter inside the compact objects must be soft.
This opens up the possibility of matter at such high density to be composed of deconfined up (u), down (d) and strange (s) quarks - the strange quark matter (SQM) \citep{1971SvA....15..347S, PhysRevD.30.272}.  In general the appearance of exotic component of matter inside the star including SQM softens the EOS supporting the maximum bound of tidal deformability obtained from two GW event observations.
This opens up the possibility of hadronic matter phase transition to de-confined SQM at the core of the compact star  giving birth to hybrid stars.
According to Bodmer and Witten conjecture \citep{bodmer1971collapsed, PhysRevD.30.272}, SQM composed of $u$ , $d$ and $s$ quarks could be true ground state of matter in strong interactions. Stability of SQM at zero external pressure and for large baryon number is demonstrated by \cite{farhi1984strange} with certain QCD parameters. With this possibility, even the whole compact object may be composed of stable SQM giving birth to strange star (SS). Within SS, with the production of strange quarks the energy per particle becomes less than that in the conventional matter with confined u-d quarks. However, the softening of matter leads to reduction in the possible maximum limit of the mass.

There are some models to study the SQM at low temperature with large baryon number. The most studied model is the original MIT bag model \citep{chodos1974new}. In this model, the hadrons are imagined as a bubble of free quarks in the vacuum and within the bag the quarks move freely but confined within it. The quarks within the bag can be considered as free Fermi gas. However, as stated above, the EOS of matter evaluated from this model is too soft to produce the observed massive compact objects. The original bag model is modified with inclusion of perturbative correction with non-zero strong coupling constant $\alpha_c$. Here, it should be mentioned that this correction is included in some \emph{ad-hoc} way to reproduce some lattice QCD result. Instead, another way of quark interaction inclusion is to introduce vector interaction between them via coupling to a vector field. If this vector interaction is repulsive, that will make the EOS stiff leading to enhancement of possible maximum mass. This model is popularly known as vector bag model, in short vBAG model and has been discussed in many works \citep{franzon2016effects, gomes2019constraining, lopes2021modified}. However, the recent observation of compact objects merger reveals that the at low density the EOS of matter should be soft. 

The proper knowledge of various astrophysical observables viz.  masses, radii, tidal deformabilities of these objects obtained from a broad spectrum of electromagnetic as well as GW events have very significant roles to play in constraining the composition and properties of dense matter inside the compact objects \citep{2007PhR...442..109L, weber2017pulsars}. Many recent observations indicate the massive compact objects near about $2~M_\odot$ \citep{2021ApJ...908L..46R, 2013Sci...340..448A}. 
Analyzing the GW170817 event data,
$\tilde{\Lambda}$ was found to be $\leq 900$ and the recent reanalysis evaluates the same to be $\leq 720$ \citep{abbott2018gw170817}.
In addition, the electromagnetic counterpart of GW170817 event (AT2017gfo) with kilonova models estimated a lower bound on $\tilde{\Lambda}$ to be $\geq 400$ \citep{Radice2018}.
The LOVE-Q relation is universal for neutron stars and quark stars as shown by \citep{2013Sci...341..365Y}. Consequently if the paritciapting stars in the binary mergers are treated to be SS, then also the bounds on $\tilde\Lambda$ and $\Lambda$  are applicable to constrain the SQM models \citep{li2021tidal,2021PhRvD.103j3010L,2021PhRvD.104h3011A}.
Based on the GW170817 event data as well as the respective electromagnetic counterparts, bounds on the maximum gravitational mass of the dense matter compact star is estimated as $M\leq 2.2$ M$_{\odot}$ \citep{2017ApJ...850L..19M, PhysRevD.96.123012, 2018ApJ...852L..25R}.
Another upper bound obtained on $\tilde{\Lambda}\leq600$ from data analysis of GW190425 event \citep{2020ApJ...892L...3A}. The analysis of GW190814 event observations suggested it to be from a binary coalescence of a $23.2^{+1.1}_{-1.0}$ M$_{\odot}$ black-hole and a $2.59^{+0.08}_{-0.09}$ M$_{\odot}$ compact object \citep{2020ApJ...896L..44A}. The nature of the secondary object involved in GW190814 event still remains unclear \citep{2020arXiv201001509B, 2020PhRvC.102f5805F, LI2020135812, 2021PhRvC.103b5808D}. At the same time, the nature of matter at lower density regime is constrained by an estimation of radius of compact stars from low-mass X-ray binaries in globular clusters. In this estimation it is found that the radius of a canonical $1.4$ M$_\odot$ {star} to be in the range $10-14$ km \citep{2018MNRAS.476..421S}. Moreover, the mass-radius measurements of PSR J$0030+0451$ \citep{2019ApJ...887L..24M, 2019ApJ...887L..21R} and recently of PSR J$0740+6620$ \citep{2021arXiv210506979M, 2021arXiv210506980R} from the NICER (Neutron star Interior Composition ExploreR) space mission also provide with adequate information regarding the radius of compact objects. The measurements of their radii also hold up to uniquely determine the dense matter EOS. The recent results from NICER observations suggest the matter inside the compact stars to be incompressible at high density regimes. 

In this work, we discuss the the observational constraints available till now in context of compact stars to constrain the SQM model parameters within interacting quark models. We consider the SQM models within bag model formalism with quark interactions and study their different aspects based on the recent observations of compact objects as described above.

The paper is arranged as follows: In sec. \ref{sec:2} we briefly discuss the SQM models we have considered in this work and aspects of tidal response. The constraints on model parameters with the numerical results are provided in sec. \ref{sec:3} and sec. \ref{sec:4} summarizes the conclusions and future perspectives. The implementation of natural units $G=\hbar=c=1$ is done throughout the work.

\section{Formalism}\label{sec:2}
\subsection{SQM models}
It is hypothesized that deconfined quark matter with $u$, $d$ and $s$ quarks are stable and true ground state of strong interacting matter. This kind of matter can be described by original MIT bag model where the quarks are considered as free to move within a bag confining the quarks. The quark interactions should be included in this naive model to make it more realistic. Here we consider the inclusion of interactions thorough two models: 1) The modified MIT bag model \citep{farhi1984strange,fraga2001small,alford2005hybrid,li2017new} and 2) Vector bag model \citep{franzon2016effects, gomes2019can, gomes2019constraining, lopes2021modified}. SQM consistently remains stable within both models at high temperature or non-zero temperature that makes these models more reliable. We consider the SQM is composed of $u$, $d$ and $s$ quarks along with electron ($e$). Here we assume $m_u=0$, $m_d=0$ and $m_s=100$ MeV for both models.
\subsubsection{Modified MIT bag model}

In this model the grand canonical potential of interacting SQM within the bag is
\begin{equation}
\label{eqn:1}
    \Omega 
    = 
    \sum_i \Omega^0_i + \frac{3}{4\pi^2}(1-a_4)\left(\mu_b\right)^4 + B_{\text{eff}}
\end{equation}
where $\Omega^0_i$ is grand canonical potential of non interacting quarks ($u$, $d$, $s$) and electrons ($e$) given by

\begin{equation}
\begin{aligned}
\label{eqn:3}
\Omega^0_q= & -\frac{{\mu_q}^4}{4\pi^2} \left( {\sqrt{1-{\eta^2_q}}\left(1-\frac{5}{2}\eta_q^2\right)} + \frac{3}{2}{\eta_q^4}\ln\frac{1+\sqrt{1-\eta_q^2}}{\eta} \right) \\  
\end{aligned} 
\end{equation}
with
\begin{equation}
 \eta_q = \frac{m_q}{\mu_q}
\end{equation}
and $q$ running for $u$, $d$ and $s$ quarks
and grand canonical potential for electron is 
\begin{equation}
\label{eqn:2}
    \Omega_e = - \frac{\mu_e^4}{12\pi^2},
\end{equation}
$B_{\text{eff}}$ is effective bag constant that includes contribution of nonperturbative QCD correction and $\mu_b$ being the baryonic chemical potential given by
\begin{equation}
\label{eqn:7}
    \mu_b = \frac{\mu_u + \mu_d + \mu_s}{3}. 
\end{equation}

Then the energy density ($\epsilon$) and pressure ($P$) can be obtained by following relations with thermodynamic potential as \citep{1996cost.book.....G}
\begin{equation}\label{eqn:8}
\begin{aligned}
\epsilon & = \Omega + \sum_{i=q,e}\mu_i n_i,  \\
P & = -\Omega,
\end{aligned}
\end{equation}
here $n_i$ denotes number density of $i$-th quark or electron that can be obtained as
\begin{equation}\label{eqn:a}
\begin{aligned}
n_i = -\frac{\partial\Omega}{\partial\mu_i}.
\end{aligned}
\end{equation}

\subsubsection{Vector bag model}

In modified MIT bag model, the QCD correction term is introduced on \emph{ad-hoc} basis to reproduce some lattice QCD data. On the other hand, in naive vBAG model the quark interaction has been included via vector-isoscalar meson field $V^{\mu}$. Further the quartic self interaction of vector interaction mediator is included to consider the Dirac sea of the quarks \citep{lopes2021modified}. The Lagrangian density of the quark matter appears similar to hadronic matter as \citep{1997NuPhA.615..441F,franzon2016effects,lopes2021modified}

\begin{equation}
\begin{aligned}
\label{eqn:9}
    {\cal L} = & \sum_{q = u,d,s}\left[\bar{\psi}_q\{\gamma^{\mu}\left(i\partial_\mu - g_{qqV}V_\mu\right)-m_q\}\psi_q-B \right] \Theta(\bar{\psi}_q\psi_q) \\ & - \frac{1}{4}{\left(\partial_{\mu}V_{\nu}-\partial_{\nu}V_{\mu}\right)}^2 + \frac{1}{2}{m^2_V}{V_\mu}V^{\mu} +  \frac{1}{4}b_4({g_{uuV}^2V_\mu}V^{\mu})^2,
\end{aligned}
\end{equation}
where $\psi_q$ and $V^\mu$ denote the field for quark $q$ and for vector interaction mediator respectively. Here $g_{qqV}$ is the vector coupling constant of quark $q$ with mediator, $m_q$ and $m_V$ are the masses of $q$ quarks and mediator respectively and $b_4$ the coefficient for self interaction.
$\Theta$ denotes heavyside function, it becomes null outside the bag and unity inside the bag.
Using mean field approximations \citep{glendenning2012compact}, chemical potential get shifted by vector field as  
\begin{align}
\label{eqn:10}
    \mu_q = \sqrt{{m^2_q}+k^2} + g_{qqV}V_0,
\end{align}
where $V_0$ is the temporal component of vector field in ground state given by the equation 
\begin{equation}
 \begin{aligned}
 \label{eqn:11}
     \left(\frac{g_{uuV}}{m_V}\right) & \sum_{q = u,d,s}\left(\frac{g_{qqV}}{m_V}\right)n_q = g_{uuV}V_0 \\ & +  \left(\frac{g_{uuV}}{m_V}\right)^2b_4(g_{uuV}V_0)^3.
\end{aligned}
\end{equation}
 The energy density and pressure are given as
\begin{align}
\label{eqn:12}
\epsilon = \sum_{q = u,d,s}(\epsilon_q) + B - \frac{1}{2}{m^2_V}{V_0}^2 - \frac{1}{4}b_4({g_{uuV}^2}V_0^2)^2  
\end{align}
and
\begin{equation}
\label{eqn:13}
    P = \sum_{q = u,d,s}(\mu_qn_q - \epsilon_q) - B + \frac{1}{2}{m^2_V}{V_0}^2 + \frac{1}{4}b_4({g_{uuV}^2}V_0^2)^2
\end{equation}
with
\begin{equation}
\begin{aligned}
\label{eqn:14}
    \epsilon_q & 
    = \frac{3}{8\pi^2}\left\{{\sqrt{{k_{f_q}}^2 + m^2}\left(2{k_{f_q}}^3 + m^2{k_{f_q}}\right)} \right. \\ & \left. + {m^4\ln{\frac{m}{\sqrt{{k_{f_q}}^2 + m^2}+k_{f_q}}}}\right\}+\frac{g_{uuV}V_0k_{f_q}^3}{\pi^2}.
\end{aligned}
\end{equation}

\subsection{Tidal deformability}

Due to extreme gravitational fields, the components in the binary compact object scenerio endure tidal deformations. To a first order approximation, this tidal deformability can be quantified by the ratio of the induced quadrupole, $Q_{ij}$ to the external static tidal field, $\varepsilon_{ij}$ as \citep{Hinderer_2008, PhysRevD.81.123016}
\begin{equation}
\begin{aligned}
\label{eqn:15}
\lambda \equiv & -\frac{Q_{ij}}{\varepsilon_{ij}} =\frac{2}{3} k_2 R^{5}, \\
\Lambda = & \frac{\lambda}{M^5}
\end{aligned}
\end{equation}
where $k_2$ is the gravitational tidal love number and $\Lambda$ is the dimensionless tidal deformability. $k_2$ is related to the compactness parameter $C\equiv M/R$ as
\begin{equation}
\label{eqn:16}
\begin{aligned}
k_2 = & \frac{8 C^5}{5}(1-2C^2)[2+2C(y-1)-y] \cdot \\ 
& \{2C [6-3y+3C(5y-8)]
 + 4C^3 [13 - 11y \\
 & + C(3y-2) + 2C^2(1+y)] + 3(1-2C^2) \cdot \\
& [2 - y + 2C(y-1)] \ln (1-2C)\}^{-1} ,
\end{aligned}
\end{equation}
where $y\equiv y(R)-4\pi R^3 \epsilon_{\text{s}}/M(R)$ with $\epsilon_{\text{s}}$ denoting the energy density at the surface of quark star and is obtained as the solution of the differential equation \citep{Hinderer_2008, PhysRevD.80.084018, PhysRevC.95.015801}
\begin{equation}
\label{eqn:17}
r \frac{dy(r)}{dr} + y(r)^2 + y(r)F(r) + r^2 Q(r) = 0,
\end{equation}
which has to be solved self-consistently with the Tolman-Oppenheimer Volkoff (TOV) equations \citep{1996cost.book.....G} with the boundary conditions $y(0)=2$, $M(0)=0$, $P(R)=0$. Here the co-efficients $F(r)$ and $Q(r)$ are defined as
\begin{equation}
\label{eqn:18}
F(r) = \frac{r - 4\pi r^3 [\varepsilon(r) - P(r)]}{r - 2M(r)},
\end{equation}
\begin{equation}
\label{eqn:19}
\begin{aligned}
Q(r) = & \frac{4\pi r [5\varepsilon(r) + 9P(r) + \frac{\varepsilon(r) + P(r)}{\partial P(r)/\partial \varepsilon(r)}]}{r - 2M(r)} \\ 
& - 4 \left[ \frac{M(r) + 4\pi r^3 P(r)}{r^2 (1 - 2M(r)/r)} \right].
\end{aligned}
\end{equation}

Analysis of the GW signal reveals another binary parameter i.e. effective tidal defomability, which is well constrained by the detectors and given by \citep{PhysRevLett.112.101101}
\begin{equation}
\label{eqn:20}
\begin{aligned}
\tilde{\Lambda} = \frac{16}{13} \frac{(12q + 1) \Lambda_1 + (12 + q) q^4 \Lambda_2}{(1 + q)^5}
\end{aligned}
\end{equation}
where $q=M_2/M_1$ is within the range $0 \leq q \leq 1$ and $M_1$, $M_2$ being the gravitational masses of the primary, secondary components involved in inspiraling BNS system.

\begin{figure*}
\begin{center}
\includegraphics[width=14cm, keepaspectratio]{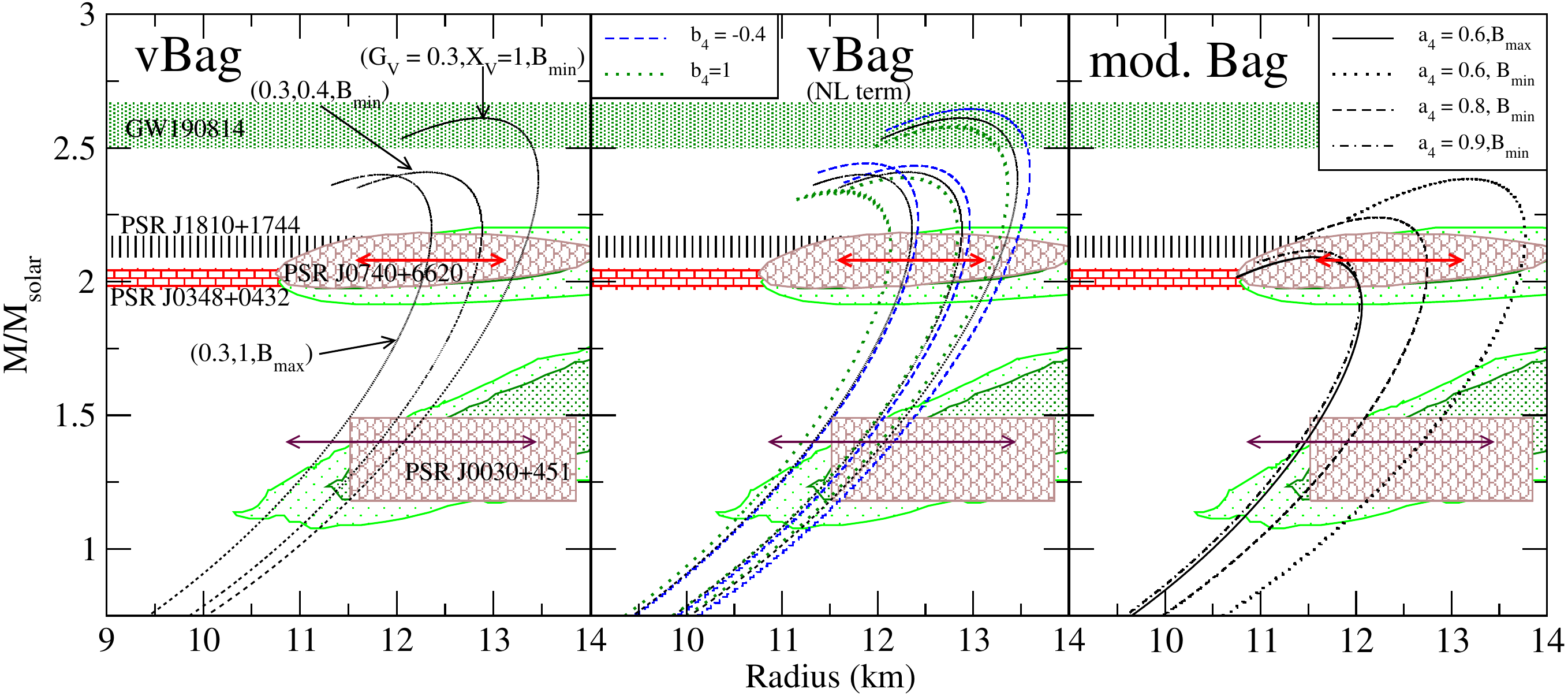}
\caption{M-R relations for different models. Left panel: with vBAG model without self interaction.
In this panel the parameter's format is ($G_V$, $X_V$, $B_{\text{eff}}$), middle panel: with self interaction in vBAG model. Right panel: modified MIT bag model.
The constraints from observations are presented in shaded regions as indicated in the figure.Three contours are for PSR J0740+6620, denser dot shaded region is for 1-$\sigma$ and rarer dot shaded region for 2-$\sigma$ 
\citep{2021arXiv210506979M}, 
bubble shape pattern contour and corresponding radius constraint is suggested by \citet{2021arXiv210506980R}.
For PSR J0030+0451 dense dot shaded region is for 1-$\sigma$ ($68\%$ credibility) and rarer dot shaded region is for 2-$\sigma$ ($95\%$ credibility) \citep{2019ApJ...887L..24M}, bubble shape pattern rectangular region is suggested by \citet{2019ApJ...887L..21R}. Horizontal straight line corresponding to $1.4$ M$_\odot$ depict radius constraints provided by \citet{2020PhRvD.101l3007L}.
Vertical small dashed lines and boxed shaded region
represent mass-radius constraints from PSR 
J1810+1744 and PSR J0348+0432 respectively
\citep{2021ApJ...908L..46R, 2013Sci...340..448A}.
}
\label{fig-001}
\end{center}
\end{figure*}

\section{Results}\label{sec:3}

In this section, we discuss the suitable parametric values for the models in context of recent astrophysical observations. First the model parameters are fixed within the stability window, {\it i.e.} the set of parameters which produces stable SQM but evaluates {\it u-d} matter with energy per particle above that in normal hadronic matter. We consider the variation of quark-vector particle interaction so that the parameter $G_V = \left(g_{uuV}/m_V\right)^2$ varies from $0.1$ to $0.3$ \citep{lopes2021modified}. In most of the literature, the coupling constant for all quarks are considered equal ($g_{uuV} = g_{ddV}=g_{ssV}$) giving the ratio $X_V=g_{ssV}/g_{uuV} = 1$.
However, symmetry group predicted the value of $X_V$ to be $0.4$ \citep{lopes2021modified}.
 Both the possibilities are considered in this work. Considering the stability window the admissible $B$ values with this chosen model parameters are shown in table \ref{tab:1}. The parameter value of self interacting non linear term ($b_4$) does not affect the stability window  \citep{lopes2021modified}. To keep the deviation from linear model minimum, the value of nonlinear parameter should be kept small as $|b_4|\leq 1$. We have opted the range of $b_4$ from $-0.4$ to $1.0$ \citep{lopes2021modified}.
Similarly, in the modified MIT bag model the \emph{ad-hoc} parameter ($a_4$) values have been tabulated for the stability window with the corresponding range of $B_{\text{eff}}$ in table \ref{tab:b}.

\begin{table} 
\caption{Maximum and minimum values of $B$ with different parameters for stable SQM with vector bag model}
\centering
\begin{threeparttable}[b]
\begin{tabular}{cccccccccccc}
\hline \hline
 $G_{V}$ & $X_V$ &Min. $B^{1/4}$ & Max. $B^{1/4}$ \\
   (fm$^2$)              & & (MeV) & (MeV) & \\
\hline
     $0.1$ & $1.0$ & 144 & 154 & \\
     $0.1$ & $0.4$ & 144 & 155 & \\
     $0.2$ & $1.0$ & 141 & 150 &  \\
     $0.2$ & $0.4$ & 141 & 152 & \\
     $0.3$ & $1.0$ & 139 & 146 & \\
     $0.3$ & $0.4$ & 139 & 150 & \\
\hline
\end{tabular}
\end{threeparttable}
\label{tab:1}
\end{table}
\begin{table} 
\caption{Maximum and minimum values of $B_{\text{eff}}$ with different parameters for stable SQM with modified MIT bag model.}
\centering
\begin{threeparttable}[b]
\begin{tabular}{cccccccccccc}
\hline \hline
 $a_4$ & Min. $B^{1/4}$ & Max. $B^{1/4}$ \\
                 & (MeV) & (MeV) & \\
\hline
    0.6 & 128 & 137 & \\
    0.7 & 133 & 144 & \\
    0.8 & 137 & 150 &  \\
    0.9 & 141 & 155 & \\
    1.0 & 145 & 159 &  \\
\hline
\end{tabular}
\end{threeparttable}
\label{tab:b}
\end{table}

First we examine the parameter space compatible with the stability window in light of the so far mass and radius of the compact objects. We have plotted the M-R curves in fig. \ref{fig-001} with the extreme boundaries of parameter space. The M-R relations have been obtained by solving TOV equations for static spherical stars. In the figure, the shaded areas represent the constraints on mass and radius obtained by analysis of data from observations of pulsars and GWs. We show the models without the non-linear term in vBAG model in the left panel, with non-linear term in middle panel and in right panel with the modified MIT bag model parameters. The maximum mass limit of a family of stars decreases with the increase of $B$ and $B_{\text{eff}}$ values. It is evident from left panel of the figure that without non-linear term, the models with $G_V=0.3$ and $X_V=1.0$, the observed lower bound on maximum mass and radius are satisfied by entire possible range of $B$. It should be noted that higher values of $B$ are admissible only when the vector interaction for $u$ quarks is near that for $s$ quarks. For lower values of $s$ quark coupling, $B$ should be less than or near to $B^{1/4}=139$ MeV. Lower value of $X_V$ decreases the maximum mass. Hence, with $X_V=0.4$, only $G_V=0.3$ with lower values B satisfy the lower bound of maximum mass. In this model, due to the repulsive nature of vector potential, the EOS is evaluated to be stiff in nature.
With decreasing value of $G_V$, the maximum mass decreases.
The lower vector potentials lead to softer EOSs  maintaining the observed M-R relation only for lower values of $B$ and equal $u$ and $s$ quark couplings. However, the curve for this set of parameters are not shown in the figure as the recent mass-radius measurement from NICER \citep{riley2021nicer} is not satisfied by less vector interaction while maximum value of $G_V$ with larger $B$ and $X_V=1.0$ marginally satisfy this constraint.
\begin{figure*} 
\begin{center}
\includegraphics[width=12cm, keepaspectratio]{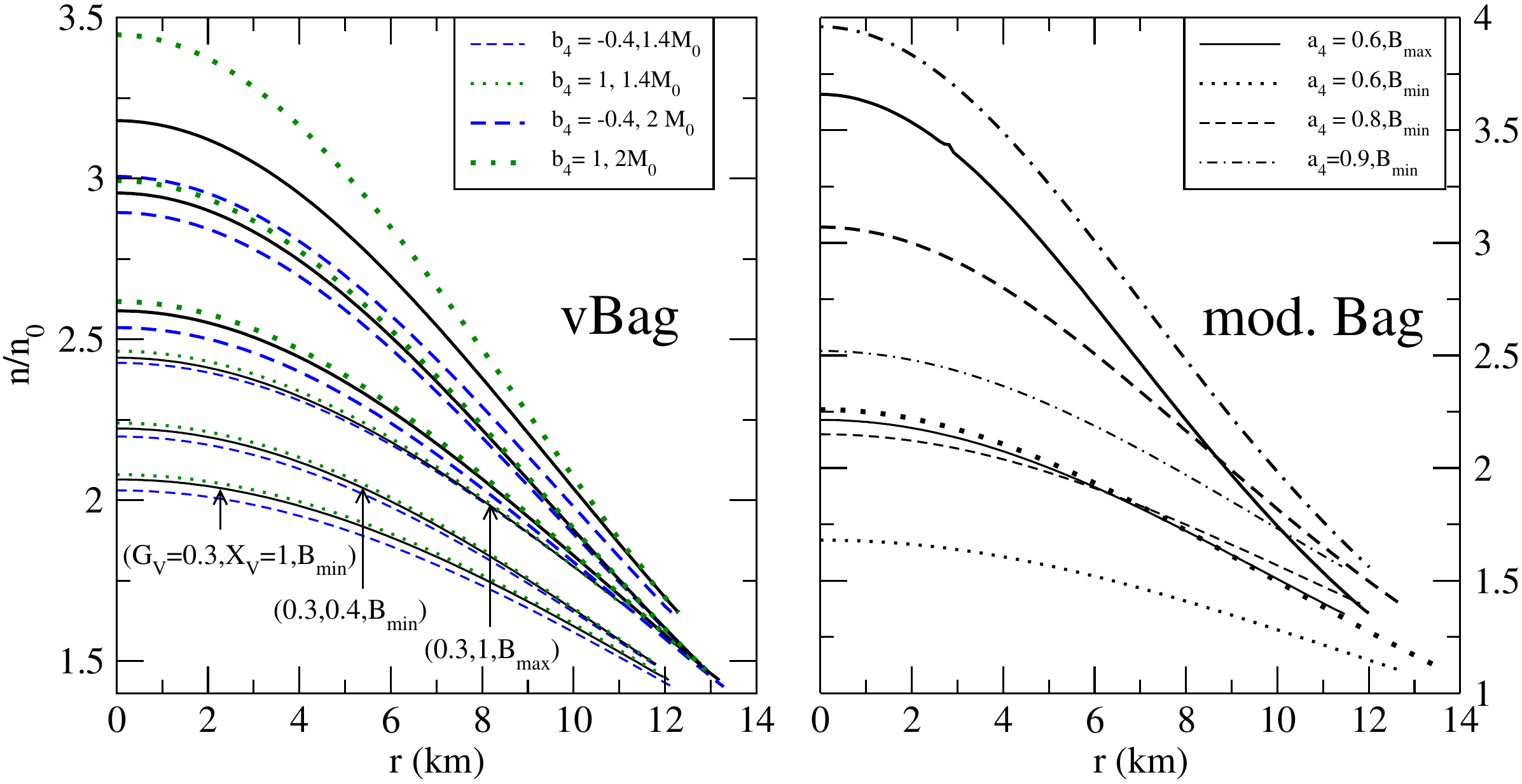}
\caption{Density profile of SSs bearing masses $1.4~M_\odot$ and $2~M_\odot$. Left panel: with vBAG model and right panel: with modified MIT Bag model. Thicker lines depict density profile for $2~M_\odot$ mass star and normal lines for $1.4~M_\odot$ mass star.}
\label{fig-a1}
\end{center}
\end{figure*}

\begin{figure*} 
\begin{center}
\includegraphics[width=14cm, keepaspectratio]{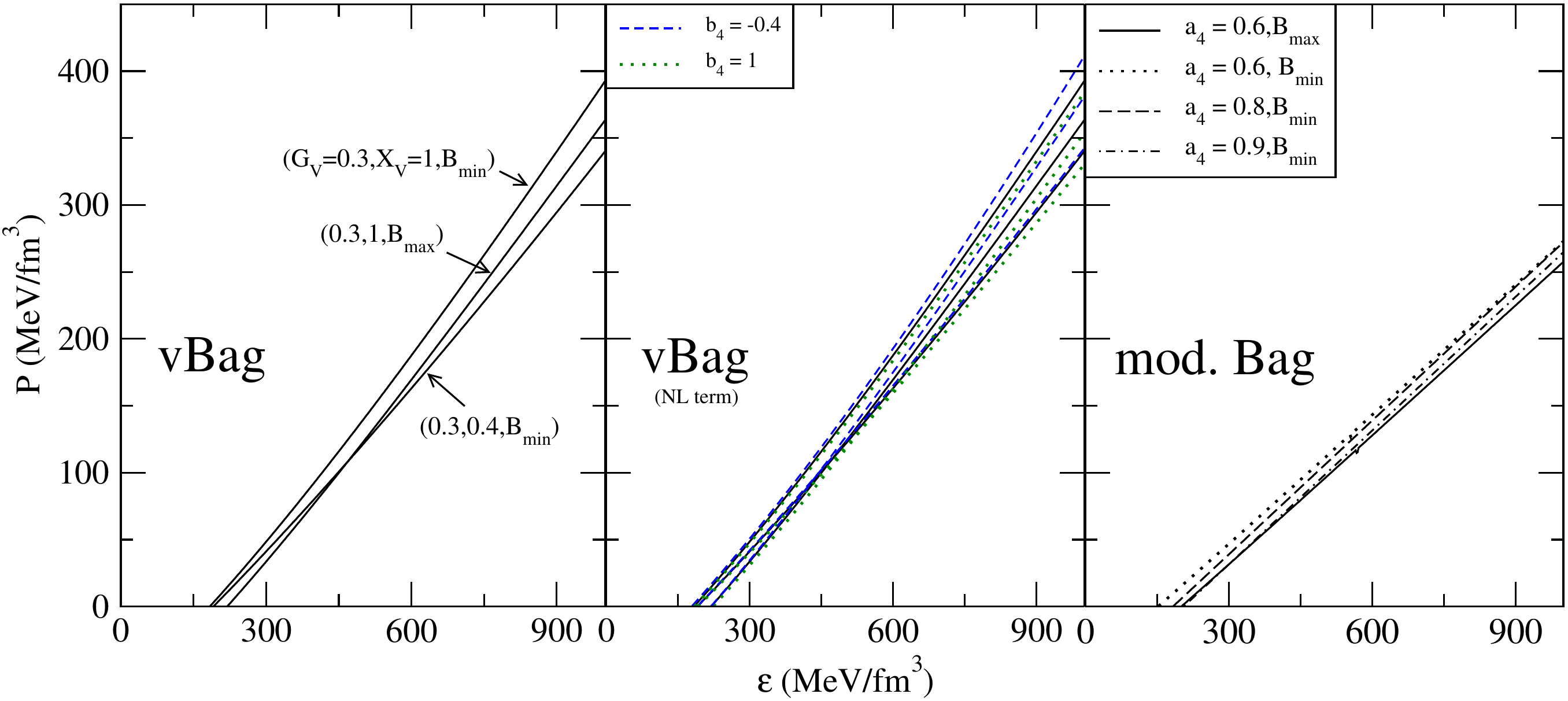}
\caption{Variation of pressure with respect to energy density (EOS) for different models. Left panel: vBAG model without self interaction, middle panel: vBAG model with self interaction term
and right panel: modified MIT bag model. The parameter set's format in all the models is similar to fig. \ref{fig-001}.} 
\label{fig-002}
\end{center}
\end{figure*}

\begin{figure*}%
    \centering
    \subfloat{\includegraphics[width=14cm,keepaspectratio ]{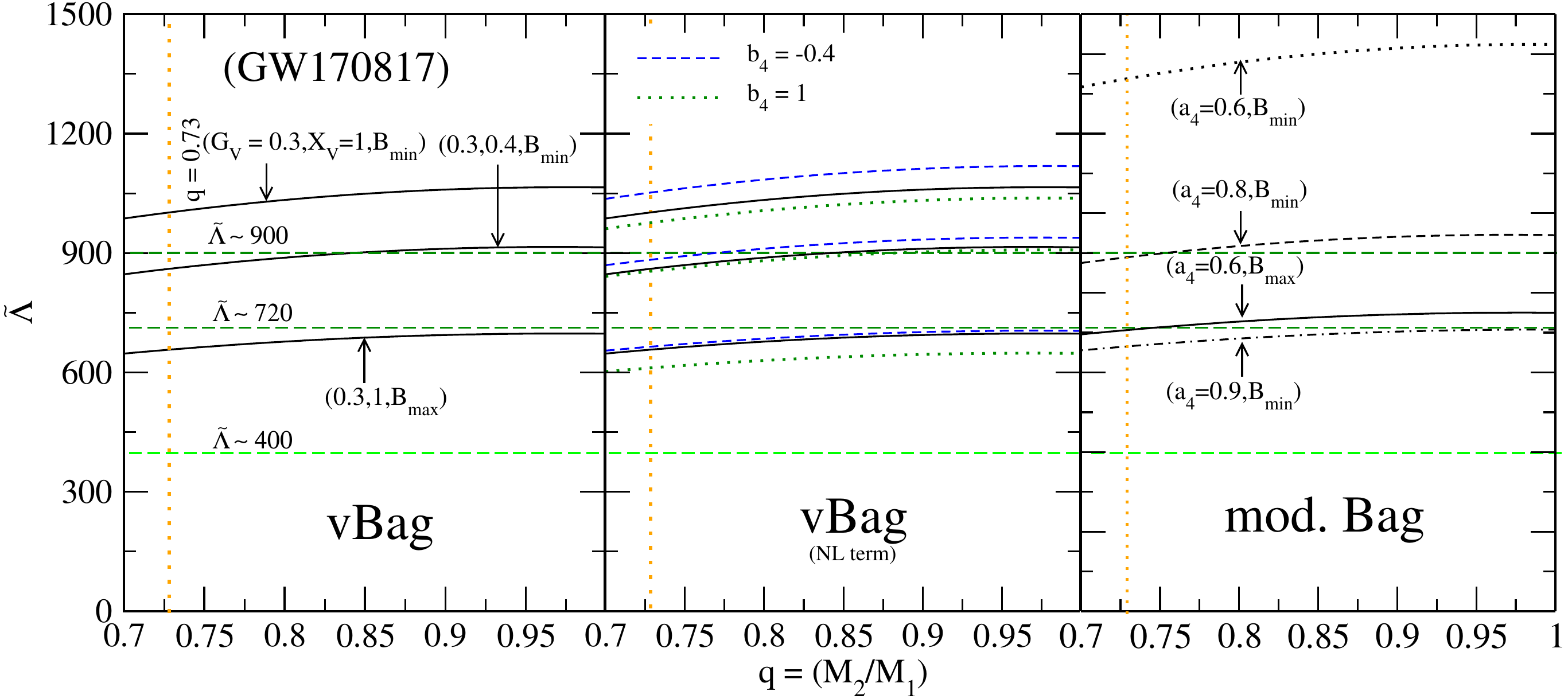}}%
    \\
    \subfloat{\includegraphics[width=14cm,keepaspectratio ]{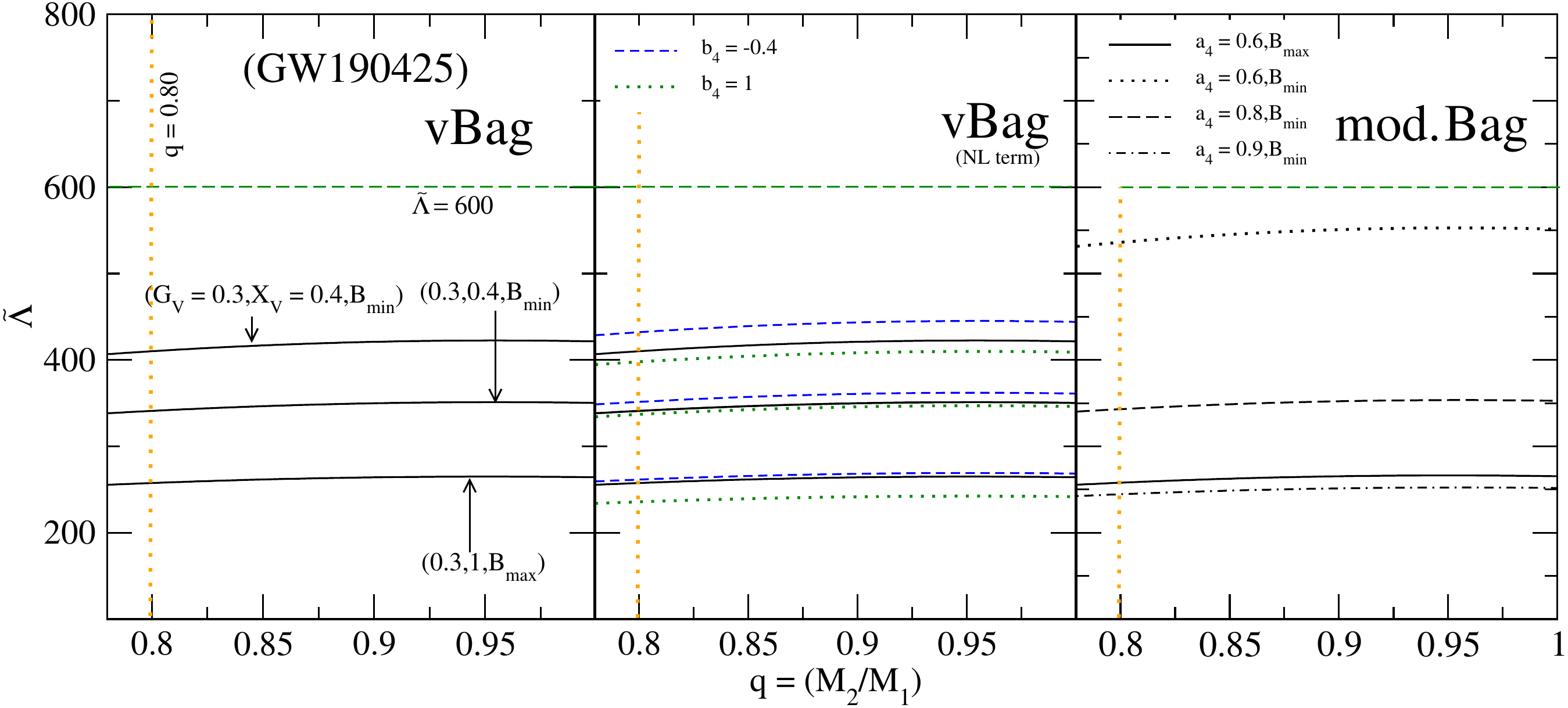}}%
    \caption{Variation of 
    $\tilde{\Lambda}$ with $q$ in
    upper panels: based on GW170817 and in lower panels:  based on GW190425 events. Left panels: vBAG model without self interaction, middle panels: with self interaction in vBAG model and right panels: modified MIT bag model. The parameter set's format in all the models is similar to fig. \ref{fig-001}. Horizontal lines are observational bound on $\tilde{\Lambda}$ and vertical dotted lines depict lower limit of of mass ratio \citep{2020ApJ...892L...3A, PhysRevX.9.011001}.
    }
    \label{fig-003}
\end{figure*}

\begin{figure*}
\begin{center}
\includegraphics[width=14cm, keepaspectratio]{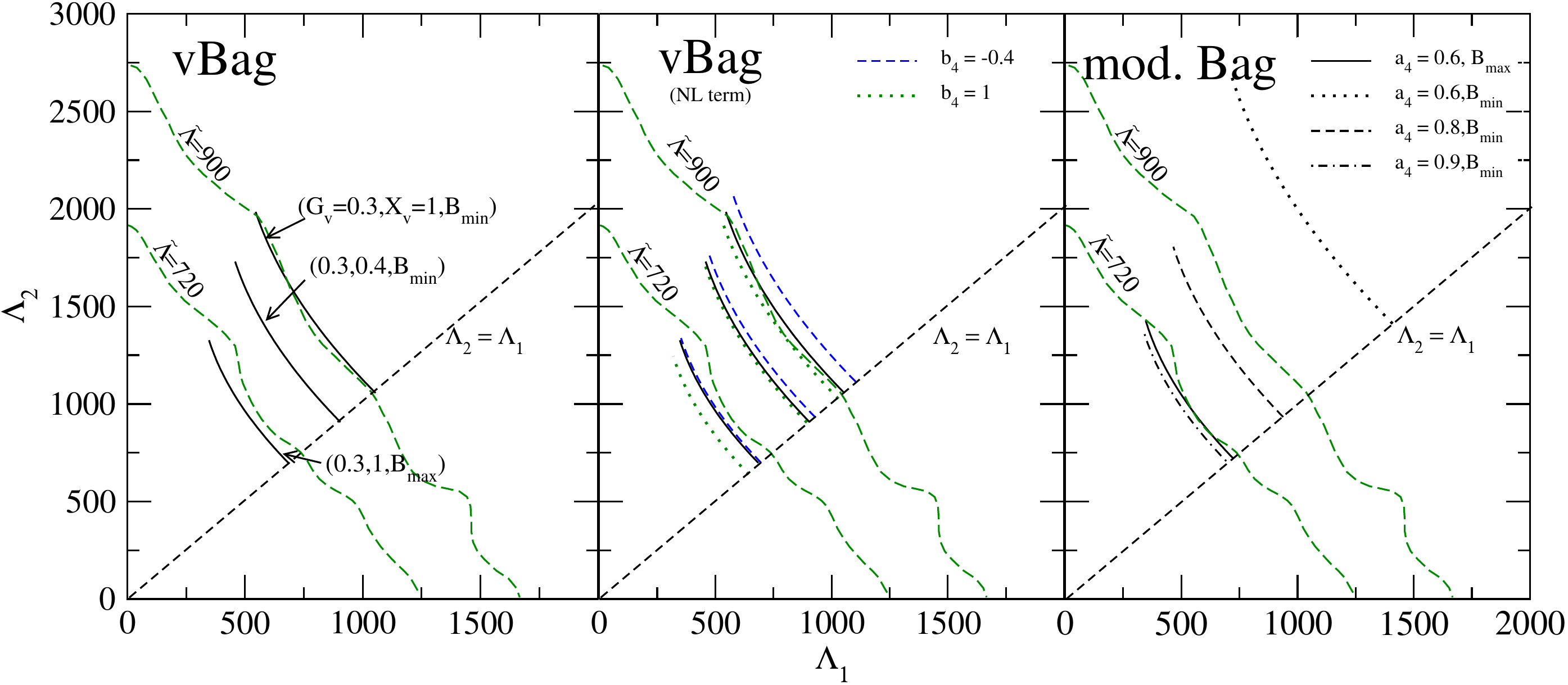}
\caption{Variation of $\Lambda_2$
with $\Lambda_1$ for binary star scenario in left panel: vBAG model without self interaction term, in middle panel: with self interaction term and in right panel: modified MIT bag model. The parameter set's format
in vBAG model is similar to fig. \ref{fig-001}.  Long-dashed curves represent upper bounds of
$\tilde{\Lambda}=900$ \citep{LIGO_Virgo2017c} (with
$90\%$ credibility) and $\tilde{\Lambda}=720$ \citep{PhysRevX.9.011001} (with
$90\%$ credibility) . Diagonal short-dashed line represents $\Lambda_2 = \Lambda_1$.}
\label{fig-004}
\end{center}
\end{figure*}

\begin{table*} 
\caption{Surface density and central density of matter with vBAG model and modified MIT bag model. Parameter's format for vector bag model is ($G_V$,$X_V$,$B_{\text{min}}$ or $B_{\text{max}}$) and for vector bag model with self interaction term is ($G_V$,$X_V$,$b_4$,$B_{\text{min}}$ or $B_{\text{max}}$).}
\centering
\begin{threeparttable}[b]
\scalebox{0.95}{
\begin{tabular}{cccc}
\hline \hline
 Model & Parameter's set & $\epsilon_s$ (MeV/fm$^3$)& $\epsilon_c$ (MeV/fm$^3$)\\
\hline
Vector & $(0.3,1.0,B_{\text{max}})$ & 222 & 978 \\
Bag & $(0.3,1.0,B_{\text{min}})$ & 184 & 796 \\
    & $(0.3,0.4,B_{\text{min}})$ & 191 & 894 \\
\\ 
Vector & $(0.3,1,-0.4,B_{\text{max}})$ &219 & 984\\
bag  & $(0.3,1,1,B_{\text{max}})$ &224 & 992\\
(NL term)  & $(0.3,1,-0.4,B_{\text{min}})$ &180 & 798\\
 & $(0.3,1,1,B_{\text{min}})$ &185&792\\ 
 & $(0.3,0.4,-0.4,B_{\text{min}})$ &190&903\\
 & $(0.3,0.4,1,B_{\text{min}})$ &192&871\\

 \\
Modified & $a_4=0.6$, B$_{\text{max}}$ & 201 & 993 \\
MIT & $a_4=0.6$, B$_{\text{min}}$ & 153 & 761 \\
 bag& $a_4=0.8$, B$_{\text{min}}$ & 184 & 888 \\
 & $a_4=0.9$, B$_{\text{min}}$ & 206 & 996 \\
\hline
\end{tabular}
}
\end{threeparttable}
\label{tab:a1}
\end{table*}
 
 Next we study the effect of vector self interaction on the M-R relation. The repulsive self interaction pushes the maximum mass up and also enhances the radius. In the entire range of non-linear term, the parameter sets with maximum value of $G_V$,$X_V=1$ and lower values of $B$ satisfy both the lower bound of maximum mass constraint and the mass-radius constraint from NICER observations. In addition with $X_V=0.4$, the models with larger values of $G_V$ and lower values of $B$ satisfy both the constraints. Hence both the constraints are well satisfied in the upper side of $G_V$, lower side of $B$ with both the values of $X_V$ throughout the entire range of $b_4$.

For comparison we plot M-R curves for modified MIT bag model in right panel. With the decreasing values of $a_4$, the EOS becomes stiff. With lowest value of $a_4$, with lower values of  $B_{\text{eff}}$ the EOS is so stiff that the radius constraints for the PSR J0740+6620 is not satisfied as suggested by \cite{2021arXiv210506980R}. However with higher values of $B_{\text{eff}}$ with minimum $a_4$, the EOS satisfies all the astrophysical mass-radius constraints. On the other hand, with increased $a_4$ as $0.8$ and $0.9$, the lower limit of star maximum mass is only satisfied with lower values of $B_{\text{eff}}$.

Here, it should be noted that the lower bound of maximum mass obtained from the mass limit of the secondary component of the GW190814 binary is satisfied by vBAG model only with $G_V = 0.3, X_V = 1$ and lower values of $B$.

The density profile inside the stars of mass $1.4~M_\odot$ and $2~M_\odot$ are shown in fig. \ref{fig-a1} for different models of SQM with different parametrizations satisfying the mass-radius constraints.
The left panel depicts density profile with vBAG model. Smaller values of u and s quark interaction ($X_V$) and higher values of $B$ produce denser SQM matter.
Density of matter increases with positive self interacting term while it decreases with negative self interacting term.
In right panel for modified MIT bag model, we observe that the density of matter increases with larger values of $a_4$ and higher values of $B_{\text{eff}}$.

The effect of different parameter values on the dense matter EOS has been shown in fig. \ref{fig-002}. The left side, represents EOSs with vBAG model. Pressure increases monotonically as function of energy density with a slight variation of slope. EOS becomes stiffer with increase in $G_V$ and $X_V$. The middle panel, represents EOSs with incorporation of self interaction term in vBAG model. Repulsive self interaction term results in stiffer EOSs. This effect is already observed from M-R relations as shown in fig. \ref{fig-001}. In right panel, pressure increases linearly with energy density for modified MIT bag model. 

\begin{figure} 
\begin{center}
\includegraphics[width=8.5cm, keepaspectratio]{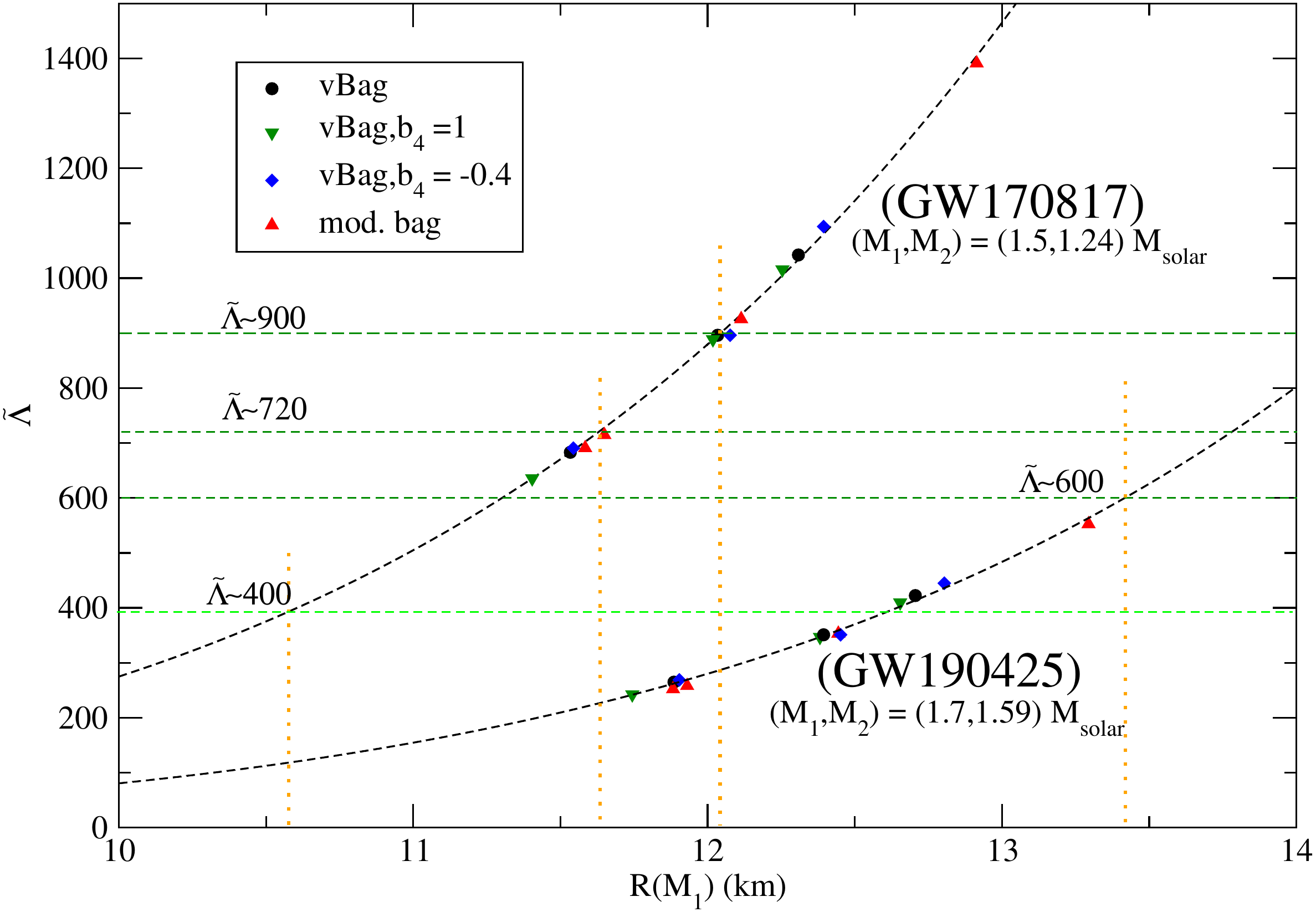}
\caption{Variation of effective tidal deformability $\tilde\Lambda$ with respect to radius of primary star $R(M_1)$. The different symbols represent different models as labelled in the figure. Horizontal straight dashed lines represent constraints on $\tilde\Lambda$ corresponding to GW170817 and GW190425 events. Vertical dotted lines denote $R(M_1)$ values corresponding to points of intersection with fitting curve.} 
\label{fig-009}
\end{center}
\end{figure}

\begin{figure*} 
\begin{center}
\includegraphics[width=15cm, keepaspectratio]{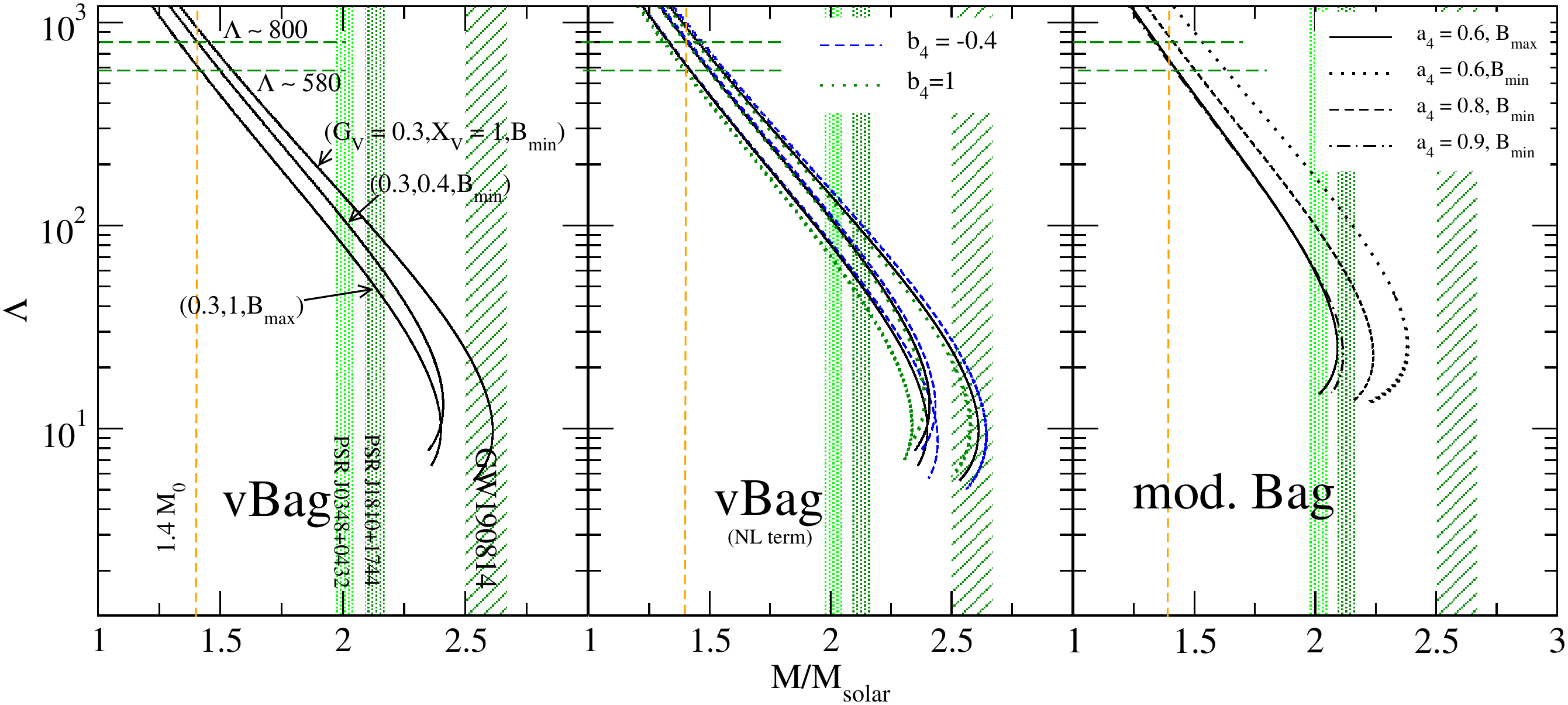}
\caption{Tidal deformability ($\Lambda$) as a function of stellar mass (in terms of solar mass) for different models. The parameter set's format in all the models is similar to fig. \ref{fig-001}. In left panel: vBAG model without self interaction, in middle panel: vBAG model with self interaction and in right panel: modified MIT bag model. Horizontal straight dashed lines represent upper bounds on $\Lambda$ \citep{abbott2017gw170817,abbott2018gw170817}. Vertical straight dashed line belongs $1.4$ solar mass. Shaded regions depict constraints on mass from NICER observation of PSR J1810+1744, PSR J0348+0432 and from GW190814 event.}
\label{fig-06}
\end{center}
\end{figure*}

Now we come to the constraints from GW observations which provide with the information about $\tilde{\Lambda}$ of a binary coalescing star pairs.
We can calculate $\tilde{\Lambda}$ in binary star scenario by solving equation \eqref{eqn:17} parallel to TOV equations and implementing eq. \eqref{eqn:20} as we discussed in previous section.
To evaluate the effective tidal deformability, we consider surface density values as provided in table \ref{tab:a1} with different quark matter EOS parameters. 
In upper panels of fig. \ref{fig-003}, we represent the variation of $\tilde{\Lambda}$ with the mass ratio ($q=M_2/M_1$) based on GW170817 event for chirp mass $\mathcal{M}=({M_1}{M_2})^{3/5}/(M_T)^{1/5}=1.188~M_\odot$ with $M_T$ being the total mass $M_1+M_2$ in the range $2.73-2.78~M_\odot$. Mass ratio of secondary to primary star $q$ varies from $0.7$ to $1$. As masses of both stars come closer, $\tilde{\Lambda}$ slightly rises. The upper bound of $\tilde{\Lambda}$ is $900$ from TaylorF2 model \citep{LIGO_Virgo2017c} and $720$ from PhenomPNRT model \citep{PhysRevX.9.011001}. We have seen that vBAG model with higher $G_V$ and lower $B$  satisfy the mass-radius relation. However, with this parameters set for equal $u$ and $s$ quark interaction, the EOS is too stiff to satisfy the upper bound of $\tilde\Lambda$. This is evident from the left and middle panel of the fig. \ref{fig-003}. Hence, this parameter set is excluded from the possible parameter combination though this set is the only set to satisfy the lower bound of mass obtained from GW190814. However, if we decrease the $s$ quark coupling constant with $X_V=0.4$, then with the lower values of $B$ the upper bound of $\tilde{\Lambda} \lesssim 900$ is satisfied up to $q=0.8$. If we increase the value of $B$ then both the upper bounds of $\tilde{\Lambda}$ are satisfied even with equal coupling of $u$ and $s$ quarks and it is obvious that with less values of $X_V$ the upper bounds are satisfied. Now if we decrease the vector interaction  both the upper and lower bounds of $\tilde{\Lambda}$ are satisfied though with less vector interaction the observed mass-radius relations are not satisfied. The inclusion of self interaction does not alter substantially the values of $\tilde{\Lambda}$. Self interaction with $b_4=-0.4$ makes the EOS stiffer and with $b_4=1.0$ makes the EOS softer. As the upper bound is satisfied marginally for $G_V=0.3$ and $B_{\rm{min}}$ with $X_V=0.4$, inclusion of nonlinear term with $b_4=-0.4$ shifts it further outside the upper bound $\tilde{\Lambda}\sim900$ near $q=1$. The result for modified MIT bag model is shown in the right panel for comparison. The matter with $a_4=0.6$ and lower values of $B_{\text{eff}}$ is too stiff to come under the upper limit of $\tilde{\Lambda}$ where as with higher values of $B_{\text{eff}}$ the matter satisfies the upper bound of $\tilde{\Lambda}\sim720$ only upto $q=0.73$. For lower values of $B_{\text{eff}}$ corresponding to $a_4=0.8$, the values of $\tilde{\Lambda}$ lie within the limits $\tilde{\Lambda}\sim900$ upto $q=0.73$ and for $a_4=0.9$ the values of $\tilde{\Lambda}$ lie within both the upper limits with higher values of $B_{\text{eff}}$.
Similarly in lower panels of fig. \ref{fig-003}, we represent $\tilde{\Lambda}$ as a function of $q$ (similar to upper panels of fig. \ref{fig-003}) based on GW190425 event. Here we consider chirp mass $\mathcal{M}=1.43M_\odot$, $M_1$ in range $1.60-1.87M_\odot$ and $M_2$ in range $1.46-1.69M_\odot$. Similar to upper panels fig. \ref{fig-003}, $\tilde{\Lambda}$ is almost independent of variation in mass ratio $q$. All $\tilde{\Lambda}$ values lie within the upper limit $\tilde{\Lambda}\sim600$ in all three panels. In other words, GW190425 event's constraint does not interrupt with validity of parameters within stability window.

\begin{figure} 
\begin{center}
\includegraphics[width=8.5cm, keepaspectratio]{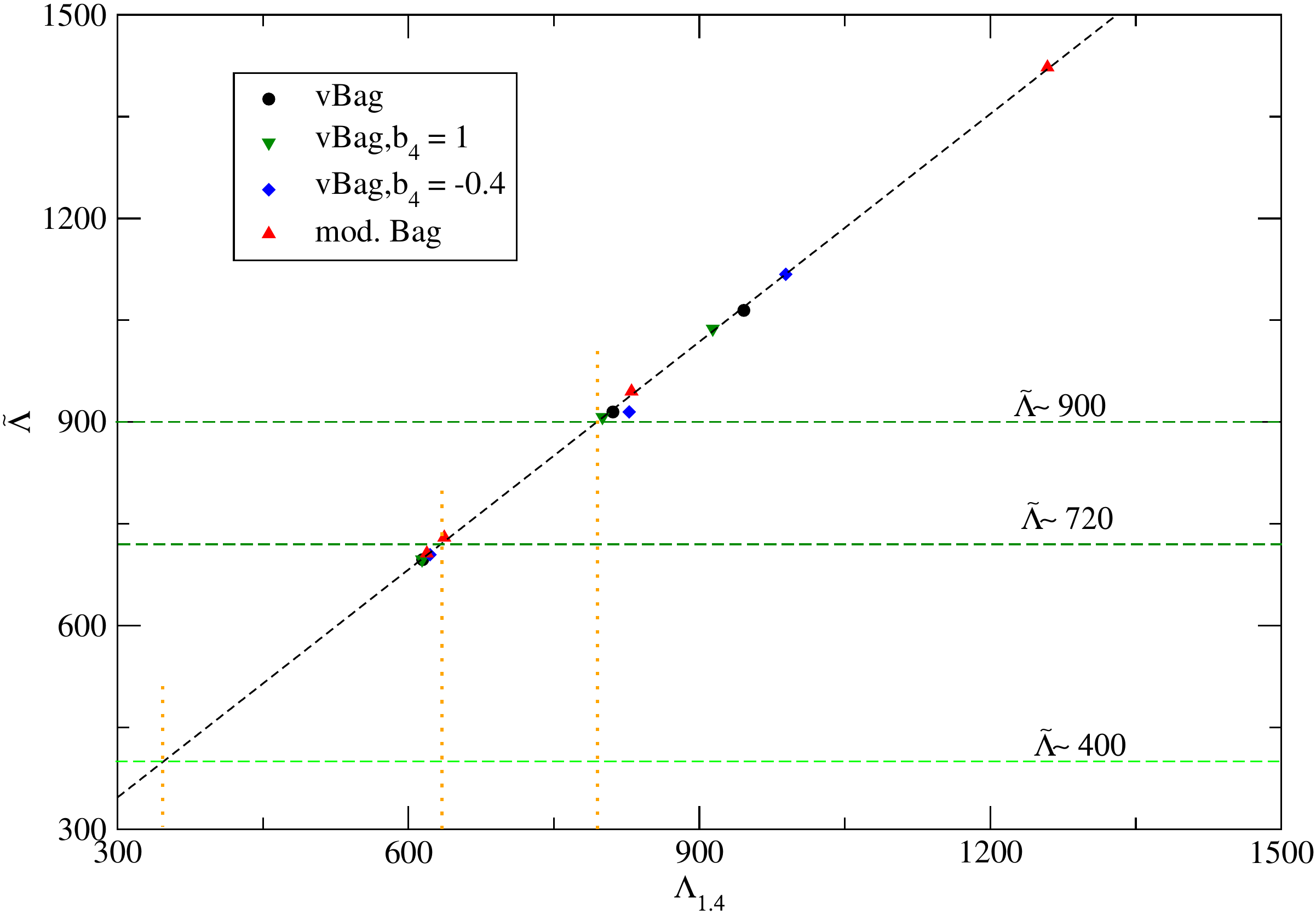}
\caption{Effective tidal deformability ($\tilde\Lambda$) of binary SS system (total mass {$M_T=2.73M_\odot$}) with respect to tidal deformability ($\Lambda$) of primary component ($1.4M_\odot$). The different symbols represent different models as labelled in the figure. Horizontal dashed straight lines correspond to upper and lower bound limits corresponding to GW170817 event. Vertical dotted lines depict $\Lambda_{1.4}$ values for points of intersection of correlation curve with upper and lower bounds.} 
\label{fig-007}
\end{center}
\end{figure}

\begin{figure} 
\begin{center}
\includegraphics[width=8.5cm, keepaspectratio]{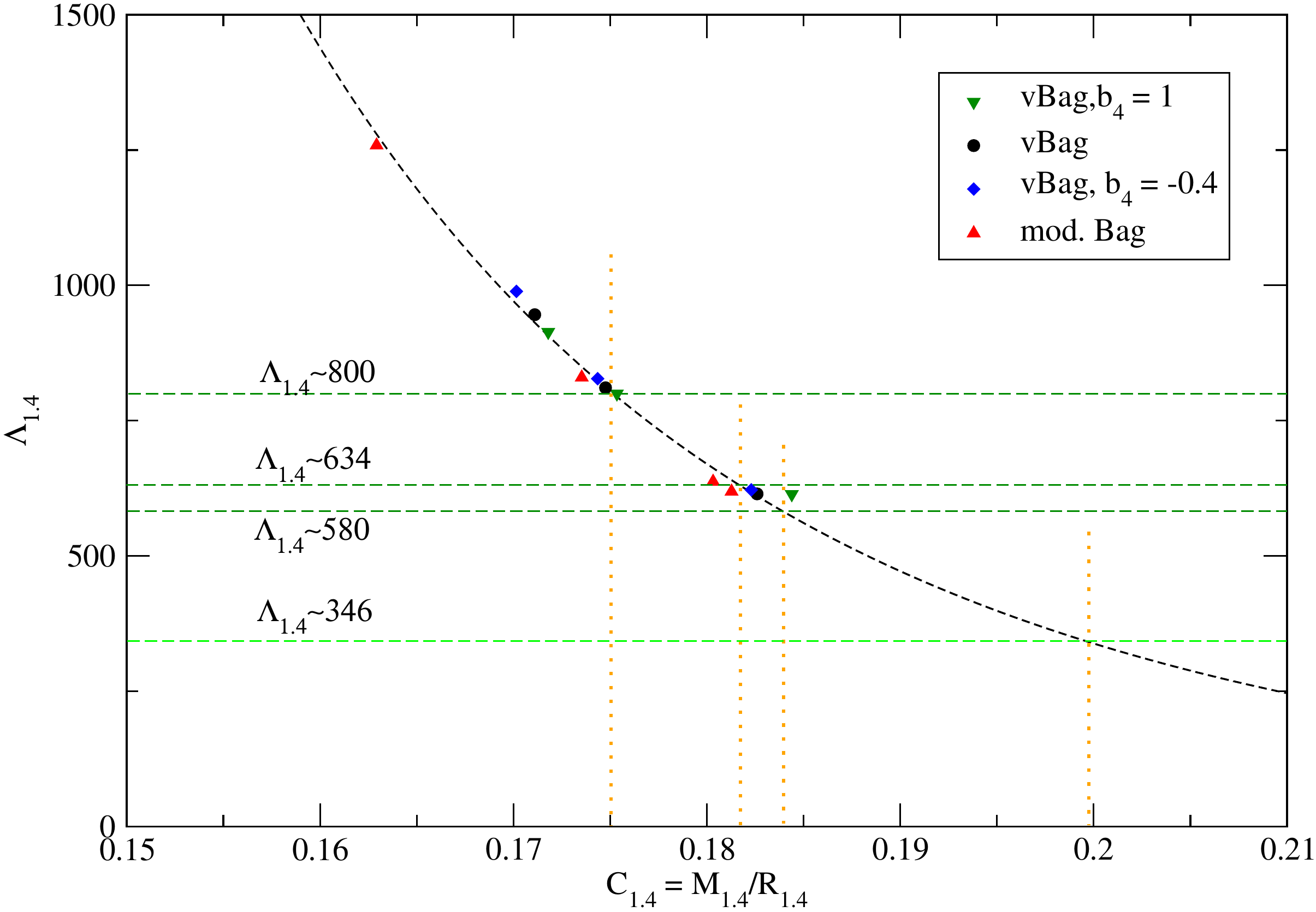}
\caption{Tidal deformability of $1.4M_\odot$ star $(\Lambda_{1.4})$ with respect to it's compactness $(C_{1.4})$. Similar to previous results, the different symbols represent different models as labelled in the figure. Horizontal lines depict upper bounds and lower bound on ($\Lambda_{1.4}$). Dashed curve represent best fit for these data points. Vertical lines are denoting $C_{1.4}$ values correspond to intersection points with correlation curve.}
\label{fig-008}
\end{center}
\end{figure}

\begin{table*} 
\caption{Tidal deformation parameters and mass-radius values with vBAG model and modified MIT bag model. Parameter's format for vector bag model is ($G_V$,$X_V$,$B_{\text{min}}$ or $B_{\text{max}}$) and for vector bag model with self interaction term is ($G_V$,$X_V$,$b_4$,$B_{\text{min}}$ or $B_{\text{max}}$). Y means satisfying constraint and N means not satisfying the constraint.}
\centering
\begin{threeparttable}[b]
\scalebox{0.95}{
\begin{tabular}{cccccccccccc}
\hline \hline
 Model & Parameter's set & $M_{\text{max}}$& $R$ & $\Lambda_{1.4}$ & $R_{1.4}$ & $C_{1.4}$ & $\tilde\Lambda$ ($q=0.73$) & PSR J$0030$ & PSR J$0740$ & PSR J$1810$ & PSR J$0348$\\
 & & ($M_{\odot}$) & (km) & & (km) & & (GW170817) & $+0451$ & $+6620$ & $+1744$ & $+432$ \\
\hline
Vector & $(0.3,1.0,B_{\text{max}})$ & 2.40 & 11.85 & 614 & 11.33 & 0.182 &658 & Y & Y & Y & Y\\
Bag & $(0.3,1,B_{\text{min}})$ & 2.61 & 12.87 & 945 & 12.07 & 0.171 & 1003 & Y & Y & Y & Y\\
    &$(0.3,0.4,B_{\text{min}})$ & 2.41 & 12.33 & 810 & 11.81 & 0.174 &861 & Y & Y & Y & Y\\
\\ 
       
Vector & $(0.3,1,-0.4,B_{\text{max}})$ &2.44 & 11.96 &622&11.34&0.182&665& Y & Y & Y & Y\\
bag  & $(0.3,1,1,B_{\text{max}})$ &2.33 & 11.56 &571&11.20&0.184&612& N & Y & Y & Y\\
(NL term)   & $(0.3,1,-0.4,B_{\text{min}})$ &2.64 & 12.92 &989&12.13&0.169&1053& Y & Y & Y & Y\\
      & $(0.3,1,1,B_{\text{min}})$ &2.57 & 12.81 &914&12.03&0.172&977& Y & Y & Y & Y\\
 & $(0.3,0.4,-0.4,B_{\text{min}})$ & 2.43&12.42&829&11.86&0.174&884& Y & Y & Y & Y\\
 & $(0.3,0.4,1,B_{\text{min}})$ &2.39&12.29&797&11.81&0.175&855& Y & Y & Y & Y\\

 \\
Modified & $a_4=0.6$, B$_{\text{max}}$ & 2.09 & 11.55 & 637 & 11.47 & 0.180 & 688 & Y & Y & Y & Y\\
MIT bag & $a_4=0.6$, B$_{\text{min}}$ & 2.38 & 13.19  & 1258 & 12.68 & 0.163 & 1338 & Y  & Y & Y & Y\\
        & $a_4=0.8$, B$_{\text{min}}$ & 2.24 & 12.22  & 830 & 11.90 & 0.173 & 890 & Y  & Y & Y & Y\\
 & $a_4=0.9$, B$_{\text{min}}$ & 2.11 & 11.54 & 619 & 11.40 & 0.181 & 665 & Y & Y & Y & Y\\

\hline
\end{tabular}
}
\end{threeparttable}
\label{tab:2}
\end{table*} 

We discussed procedure of evaluation of tidal deformability ${\Lambda}$ in section \ref{sec:2}. Tidal deformabilities of both stars are not independent of each other in binary compact star scenario for a particular EOS.  Fig. \ref{fig-004} depicts tidal deformabilities of both stars assuming chirp mass  $\mathcal{M}=1.188{M_\odot}$, with $M_T$ in the range $2.73-2.78 M_\odot$. $\Lambda_1$ and $\Lambda_2$ are tidal deformabilities of primary star with $M_1$ in the range $1.36-1.60M_\odot$ and secondary star with $M_2$ in the range $1.17-1.36M_\odot$ respectively. If we alter the values of $M_1$ and $M_2$ in their range, $\Lambda_2$ decreases with increase in $\Lambda_1$. The curve for EOS with $G_V=0.3$, $X_V=0.4$ and minimum $B$ without any self interaction lies within the upper bound $\tilde{\Lambda}\sim900$ contour which is shown in the left panel of the fig. \ref{fig-004}.Inclusion of attractive self interaction term with $b_4=1.0$ (middle panel) makes it fall within the contour. On the other hand, even without nonlinear term but with higher values of $B$ the curves lie below the both upper bound curves obtained from observation. It is obvious that with less vector interaction the curve lies below the observed contour, but these parameter set does not satisfy the mass-radius relations. Consequently, best suited parameter set with both mass-radius and GW observation is with greater vector interactions but less $s$ quark interaction, with lower values of $B$ both with and without self interaction. However another upper bound $\tilde\Lambda\sim720$ excludes both the sets of parameters with lower values of $B$. For modified MIT bag model,
the curve for higher values of $B_{\text{eff}}$ and $a_4=0.6$  marginally satisfy the $\tilde{\Lambda}\sim720$ contour while with lower values of $B_{\text{eff}}$ curve lies outside both contours. For $a_4=0.9$ with minimum value of $B_{\text{eff}}$ curve lies inside both the contours. The curve for $a_4=0.8$ with lower values of $B_\text{eff}$ lies within $\tilde{\Lambda}\sim900$ contour but outside $\tilde{\Lambda}\sim720$ contour.

Now we attempt to restrict the radius of the stars participating in the binary merger with SS composed of these EOSs from the GW observations. We plot the variation of $\tilde\Lambda$ with the radius of one star of the binary in fig. \ref{fig-009} for the parameter sets which satisfy both the mass-radius and GW observations. For this plot, we take the  primary star masses to be $1.5M_\odot$ and $1.7M_\odot$ for GW170817 and GW190425 events respectively. The plot shows strong correlation as shown in the figure by
\begin{equation}
\begin{aligned}
{{{\tilde\Lambda_{\text{fit}}}^{\text{GW170817}}}=1.15\times10^{-4}(R(M_1))^{6.38}}
\end{aligned}
\end{equation}
 for GW170817 observations. In this fit, the maximum deviation is estimated to be $(|{\tilde\Lambda_{\text{fit}}}-{\tilde\Lambda}|/\tilde\Lambda)\sim1.85\%,~\text{chi-squared value}~ {\chi}^2=\sum_{i}({{\tilde\Lambda_{\text{fit}}^i}}-{\tilde\Lambda})^2/\tilde\Lambda^i\sim1.094$, coefficient of determination, ${\mathcal{R}}^2\sim0.998$. For the GW190425 observations the fit is given by
\begin{equation}
\begin{aligned}
{{{\tilde\Lambda_{\text{fit}}}^{\text{GW190425}}}=1.2\times10^{-5}(R(M_1))^{6.83}}
\end{aligned}
\end{equation}
with the maximum deviation as $ \sim4.27\%,{\chi}^2\sim1.74$ and ${\mathcal{R}}^2\sim0.994$.
Then the upper bounds on $\tilde\Lambda\leq 900$ and $\tilde\Lambda\leq 720$ from the observation of GW170817 limit the upper radius by $\lesssim 12.04$ km and $\lesssim 11.62$ km respectively. Another lower bound from the same observation $\tilde\Lambda\leq 400$ limits the lower radius by $\geq10.57$ km of $1.5$ solar mass star.
Likewise, the upper bound on $\tilde\Lambda$ for GW190425 observations provide with limit $\lesssim 13.41$ km for a star of $1.7$ solar mass.

Next we plot the variation of tidal deformability $\Lambda$ with the stellar mass in fig. \ref{fig-06}. In left panel, the set of parameters for higher interactions and lower values of $B$ do not satisfy upper bound $\Lambda_{1.4}\sim800$ even after inclusion of self interaction term as shown in middle panel. If we decrease $X_V$ to $0.4$ then it marginally satisfy this limit and inclusion of self interacting term with $b_4=-0.4$ makes it satisfy the limit. In right panel with modified MIT bag model the parameter sets $a_4=0.9$ with lower values of $B_{\text{eff}}$ and $a_4=0.6$ with higher values of $B_{\text{eff}}$ are enough soft to satisfy the values of $\Lambda$ for typical $1.4~M_\odot$ mass star. Upper bounds on $\Lambda$ of $1.4~M_\odot$ star are suggested by \cite{abbott2017gw170817, abbott2018gw170817}.
Another upper bound  $\Lambda_{1.4}\lesssim 580$ is not satisfied by the all sets of parameters with vBAG model. Only with inclusion of attractive self interaction with $B_{\rm{max}}$ and $G_V=0.3$, $X_V=1$ the bound $\Lambda_{1.4}\lesssim 580$ is satisfied. In modified MIT bag model with parameter $a_4=0.6$ and $0.9$ corresponding to higher and lower value of $B_{\text{eff}}$ respectively satisfy this limit.

The effective tidal deformability $\tilde\Lambda$ shows very tight correlation with the tidal deformabilities $\Lambda$ of individual stars. This is shown in the fig. \ref{fig-007} by the variation of $\tilde\Lambda$ with $\Lambda$ of a $1.4~M_\odot$ star. A linear correlation $\tilde\Lambda=1.119\Lambda_{1.4}+11.39$ is obtained between $\tilde\Lambda$ and $\Lambda_{1.4}$ with maximum deviation $\sim0.803\%$. ${\mathcal{R}}^2$ is approximately $0.999$ and ${\chi}^2 \sim0.189$. We obtain new upper bounds $\Lambda_{1.4}\sim800$, $\Lambda_{1.4}\sim634$ and lower bound $\Lambda_{1.4}\sim346$ on $\tilde\Lambda$ corresponding to $\tilde\Lambda\sim900$, $\tilde\Lambda\sim720$ and $\tilde\Lambda\sim400$ respectively. 

In fig. \ref{fig-008}, we plot $\Lambda$ of $1.4~M_\odot$ star with respect to it's compactness $C_{1.4}$ including newly obtained $\Lambda_{1.4}$ constraints. Behaviour of $C_{1.4}$ with different parameters can be predicted from fig. \ref{fig-001}.
The new constraints $\Lambda_{1.4}\sim634$ and $\Lambda_{1.4}\sim346$ do not make any difference in validity of parameters as predicted in fig. \ref{fig-003}. For data points corresponding to $\Lambda_{1.4}$ and $C_{1.4}$, correlation $\Lambda_{1.4}=0.0099/(C_{1.4})^{6.49}$
seems best fit. Here maximum deviation is approximately $3.94\%$ with $\chi^2=4.094$ and $\mathcal{R}^2\sim0.9927$.
We get lower limits $C_{1.4}\sim 0.175$,$C_{1.4}\sim0.182$,$C_{1.4}\sim0.184$ and $C_{1.4}\sim0.199$ corresponding to $\Lambda_{1.4}\sim800$,$\Lambda_{1.4}\sim634$,$\Lambda_{1.4}\sim580$ and $\Lambda_{1.4}\sim346$ respectively.

We summarize the astrophysical compatibility of different parametrizations considering different model in table \ref{tab:2}. We also enlist various star quantities corresponding to the respective models. Validity of various model parameters with different pulsar's observations is also shown in table \ref{tab:2}.
\section{Conclusions}\label{sec:4}
In this work, we discuss the possibility of existence of SS in light of the recent astrophysical observations within the framework of bag model. The original bag model considers the free quark contained in a bag. However, for simple MIT bag model maximum $1.85~M_\odot$ is feasible within stability window \citep{lopes2021modified}. The recent observations of massive compact stars indicate that in the model some interaction should be incorporated as the inclusion of repulsive vector interactions in simple MIT bag model provide with stiffer EOSs. Following this argument, we sort the correct parameters for bag model with vector interaction from several recent mass-radius observations of compact stars. It shows that in order fulfill the observed mass-radius bounds, the SQM EOS should be stiff with large vector interaction. However, the constraints of softness of the EOS obtained from the GW observations show that maximum interaction with higher values of $B$ can produce such stiff EOS. Hence we conclude that intermediate interaction with lower values of $B$ are correct choice of parameters of SQM. 
Further inclusion of attractive interaction term softens the EOS implying self interaction of vector particle as an open possibility. We have also discussed the possibility of inclusion of attractive and repulsive self interaction of the vector particle. In light of these constraints we can predict the admissible parameter sets. Comparing to modified MIT bag model, where the interaction term is introduced on \emph{ad-hoc} basis, it is also found that intermediate interaction with coefficient $a_4=0.9$ and $a_4=0.6$ with lower values of $B_{\text{eff}}$ and higher values of $B_{\text{eff}}$ respectively satisfy both the constraints from mass-radius and GW observations. Based on our analysis and subsequent constraints on ${\tilde\Lambda}$, we can see that secondary component of GW190814 does not seem to be a SS.

Then, after fixing the parameter sets, we try to fix the limit on primary component's radius involved in GW170817 event with mass $1.5~M_\odot$ which comes out to be ($10.57\leq R_{1.5}\leq12.04$) km with different estimation of $\tilde{\Lambda}$ upper limits depending upon different interpretations of the observed data.
In addition we also find the limit on tidal deformability $\Lambda_{1.4}$ of a typical $1.4~M_\odot$ star from the observed limit of $\tilde{\Lambda}$.
It is found that upper bound on $\Lambda_{1.4}$ lies between $\sim 634$ to $\sim 800$, while the lower
bound is $\sim 346$.

The new limit $\Lambda_{1.4}\sim800$ corresponding to $\tilde\Lambda\sim900$ tallies with the upper bound suggested by \cite{abbott2017gw170817}. The other maximum limit $\tilde\Lambda\sim634$ is showing $9.3\%$ deviation from the upper bound obtained through reanalysis of GW170817 data \citep{abbott2018gw170817}. We also obtain range $0.175\leq C_{1.4}\leq0.199$ on compactness parameter for a SS of $1.4~M_\odot$. Conclusively, tidal deformation parameters for SQM with vBAG model and modified MIT bag are good match with GW170817 observation with intermediate interaction.

The concept of hybrid star is another vital inference from recent astrophysical observations and studied in various recent works \citep{Nandi_2018, 2019ApJ...877..139G, 2018PhRvD..97h4038P, 2019MNRAS.489.4261M, 2021EPJST.tmp....5N, PhysRevC.103.055814} as already mentioned in sec. \ref{sec:1}. Further analysis on the aspect of hybrid star configurations is beyond the scope of this work and will be addressed in future studies.

\section*{Acknowledgements}
The authors thank the anonymous referee for constructive comments that significantly contributed to enhancing the manuscript's quality.
VBT is thankful to Debades Bandyopadhyay for fruitful discussions.

\section*{Data Availability}
Data sharing not applicable to this article as no data sets were generated during this study.



\bibliographystyle{mnras}
\bibliography{example} 



\bsp	
\label{lastpage}
\end{document}